\documentclass[preprint2]{aastex63}

\received{}
\revised{}
\accepted{}
\shorttitle{On the Paradoxical Impact of Blending by Red Clump Giants}
\shortauthors{}
\graphicspath{{./}{figures/}}

\begin{document}

\title{On the Paradoxical Impact of Blending by Red Clump Giants}

\correspondingauthor{Daniel Majaess}
\email{daniel.majaess@msvu.ca}

\author{Daniel Majaess}
\affiliation{Mount Saint Vincent University, Halifax, Nova Scotia, Canada.}
\affiliation{Saint Mary's University, Halifax, Nova Scotia, Canada.}



\begin{abstract}
The impact of blending by RCGs (red clump giants, or relatively metal-rich red horizontal branch stars) is discussed as it relates to RRab and classical Cepheids, and invariably establishing an improved distance scale.  An analysis of OGLE Magellanic Cloud variables reaffirms that blending with RCGs may advantageously thrust remote extragalactic stars into the range of detectability.  Specifically, simulations of Magellanic Cloud RRab and RCG blends partly reproduce bright non-canonical trends readily observed in amplitude-magnitude space ($I_c$ vs.~$A_{I_c}$).  Conversely, the larger magnitude offset between classical Cepheids and RCGs causes the latter's influence to be challenging to address.   The relative invariance of a Wesenheit function's slope to metallicity (e.g., $W_{VI_c}$) implies that a deviation from the trend could reveal blending and photometric inaccuracies (e.g., standardization), as blending by RCGs (a proxy of an evolved red stellar demographic) can flatten period-Wesenehit relations owing to the increased impact on less-luminous shorter-period Cepheids.  That could partly explain both a shallower inferred Wesenheit function and over-estimated $H_0$ values. A consensus framework to identify and exploit blending is desirable, as presently $H_0$ estimates from diverse teams are unwittingly leveraged without homogenizing the disparate approaches (e.g., no blending correction to a sizable $\simeq 0.^{\rm m}3$).  
\end{abstract}


\section{Introduction}
Blending has been identified in ground-based observations of standard candles in globular clusters, the Galactic Bulge, galaxies, and high-resolution HST data \citep[e.g.,][]{ma12,ri16}. Blending is defined here as arising from unresolved stars along the sightline falling within a standard candle's PSF (e.g., classical Cepheid), yet which remain unaccounted for when extracting photometry. The extraneous flux can result in underestimated distances, and may be challenging to address depending partly on the magnitude offset between the target (e.g., RRab or classical Cepheid) and coincident star(s) (e.g., RCGs, red clump giants or comparatively metal-rich red horizontal branch stars).  The blends investigated here arise from chance superpositions with RCGs, which are abundant in the solar neigborhood and beyond, as confirmed observationally and by stellar models.  As a result the stars are employed to trace Galactic structure, characterize extinction laws, and establish stellar cluster and galaxy distances \citep[][and references therein]{ni05,gr07}. As shown here, RRab and RCG blends can be identified in the amplitude-magnitude plane (e.g., $I_c$ vs.~$A_{I_c}$), and paradoxically photometric contamination may be exploited as it can advantageously propel faint extragalactic variables into the detection threshold \citep{ma18}.   A key impetus of the present work is to support that assertion through simulations of RCGs grafted upon OGLE observations of variables in the Magellanic Clouds. 

Blending's impact on remote classical Cepheids is debated (e.g., null correction to $\simeq 0.^{\rm m}3$), with concerns emerging near the conclusion of the HST project to secure $H_0$ as the additional flux can lead to an overestimated expansion rate.  For reviews and rebuttals see \S$8.5$ in \citet{fr01}, \S$7$ in \citet{mo00}, and \S$8$ in \citet{mo01}.  Since that era additional data provide an enhanced understanding of the important degeneracies between blending, characterizing the effect of metallicity on classical Cepheid distances, and non-standard extinction laws.  A method to infer the impact of metallicity is to examine changes in the Wesenheit relation as a function of galactocentric distance, and thus abundance \citep[][their Fig.~1]{lu11}.  However, a problem arises owing to concurrent gradients in stellar density and surface brightness. Importantly, \citet{su99} and \citet{ma01} stressed that contamination by neighboring stars near the crowded central region of a galaxy (e.g., M101) may bias the flux of more metal-rich classical Cepheids, and could compromise determinations of the metallicity effect. Assessing whether chemical composition affects the slope ($\alpha$) of the Wesenheit ($W_{VI_c}$) function is critical for constraining blending.   A  degeneracy occurs since blending can preferentially impact shorter-period Cepheids (relative flux). In sum, the slope of the Wesenheit function may be a pertinent means for identifying blending if a given relation is insensitive to metallicity (slope \& magnitude, and the latter is discussed in \S \ref{s-cc}).  However, \citet[][their Fig.~12]{ri09a} implied that the slope of the Wesenheit ($W_{VI_c}$) function is sensitive to abundance and for metal-rich classical Cepheids $\alpha \simeq -3.0$, as inferred from numerous variables spanning throughout SN-host galaxies such as NGC1309 and NGC3021. \citet{ma11} countered that the $W_{VI_c}$ slope is comparatively constant across a sizable metallicity baseline ($\alpha \simeq -3.3$, $\Delta$[$\rm{Fe/H}$]$\simeq 1$, their Fig.~1), as established from analyzing relatively local classical Cepheids in the LMC, SMC, Milky Way, NGC6822, and IC1613, and \citet{ma10} cited that the deviant $\alpha$-results determined by \citet{ri09a} and \citet[][NGC5128]{fe07} arose from potentially inaccurate photometry or blending corrections. \citet{ma10} likewise relayed that the \citet{ri09a} classical Cepheids for NGC1309 and NGC3021 were too blue ($V-I_c$), and yielded a nearly negligible or negative mean reddening. Reassessments of photometry are essential. \citet[][S$H_0$ES]{ho16} updated the team's earlier analysis, and conveyed in their companion study \citep{ri16} that a global $W_{VI_c}$ slope of $\alpha=-3.38\pm0.02$ was determined, therefore presumably overturning the prior overarching hypothesis \citep[][their Fig.~12]{ri09a}. The lack of consensus on this topic is further discussed in \S \ref{s-cc}.  

\begin{figure}
\begin{center}
\includegraphics[width=8cm]{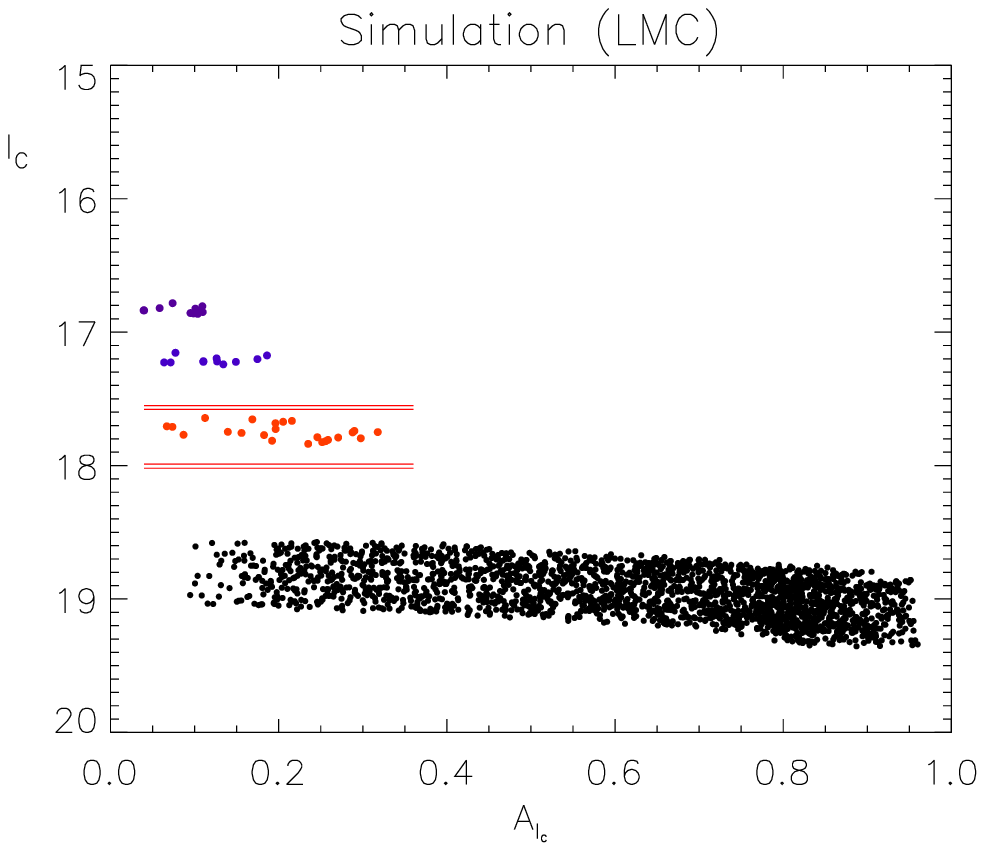} 
\includegraphics[width=8cm]{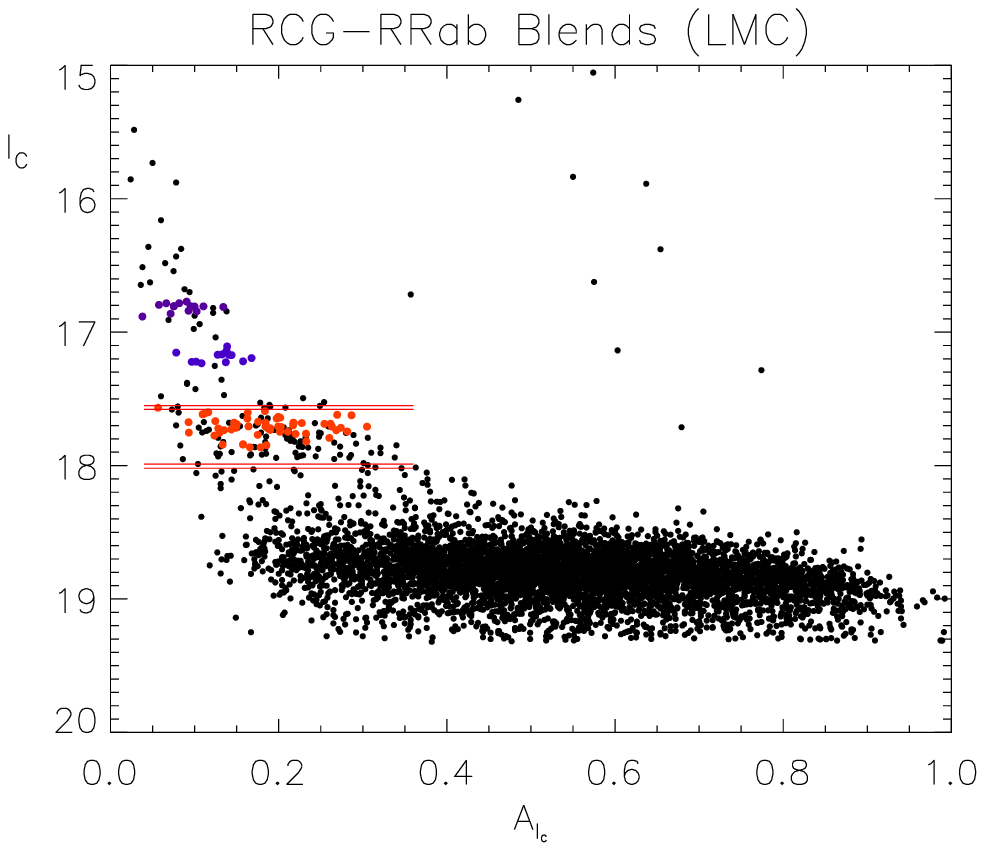} 
\includegraphics[width=8cm]{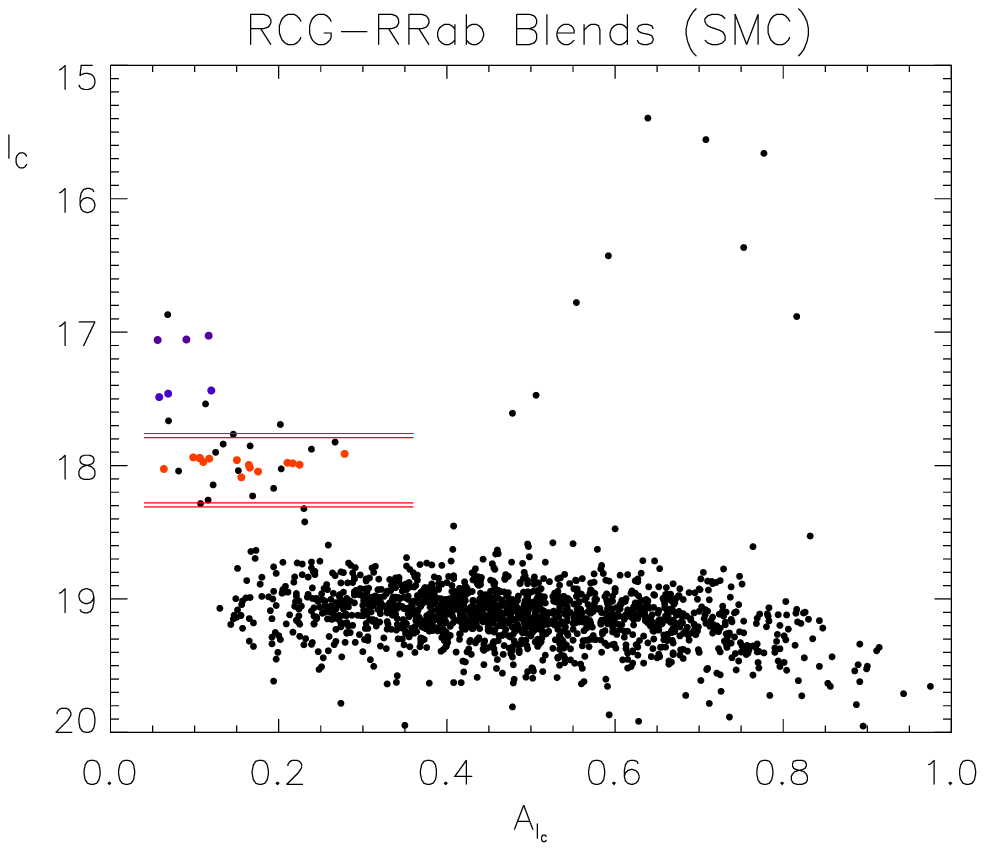} 
\caption{Blends of RRab and RCG stars (open circles) simulated from first principles (top), and grafted upon OGLE observations of the Magellanic Clouds (bottom panels).  The blended stars are readily discernible in the magnitude ($I_c$)-amplitude ($A_{I_c}$) plane owing to the relatively comparable magnitudes between RRab and RCG stars, and the abundant nature of the latter. Importantly, the overdensity near $I_c \simeq 17^{\rm m}.8$ is reproduced from blending by a single RCG ($n=1$), and the brighter ridges represent $n=2,3$.}
\label{fig-l}
\end{center}
\end{figure}

This study was inspired by the aforementioned context, in tandem with desiring to confirm that blending with ubiquitous RCGs can thrust extragalactic variables into the realm of detection.  In \S \ref{s-rrab} it is demonstrated that observed trends tied to LMC RRab can be reproduced in part by including contaminating flux from RCG stars (particularly the $n=1$ ridge).  The minimal magnitude offset between the classes allows the impact of blending to be identified.  That is not necessarily true for classical Cepheids (\S \ref{s-cc}), and the trend is comparatively challenging to discern at remote distances owing to the uncertainties, and the analysis shifts to analyzing deviations in the slope of the Wesenheit function (e.g., a potential indicator of blending).  

\section{Analysis}
\subsection{RRab \& RCG blends}
\label{s-rrab}
Blended RRab and RCG stars were first simulated by adopting RRab period-magnitude-amplitude relations and adding scatter (Fig.~\ref{fig-l}, top).  Subsequent to that, the RRab and RCG blends were better approximated by grafting the RCG magnitude ($I_{RC}$) directly onto OGLE Magellanic Cloud RRab observations \citep{so09,so10b}.   A mean LMC RCG magnitude of $I_{RC}\simeq18.2$ was adopted \citep{al02}, and the revised blended magnitude ($I_{B}$) was calculated via:
$I_{B}\simeq-2.5$ $\log{(10^{-I_0/2.5}+\sum_{}^{} n 10^{-I_{RC}/2.5})}$, where $n$ defines the number of RCG blends.  The blended RRab amplitudes ($A_{B}$) were estimated by determining the offset between contamination added at maximum ($-0.5 A_0$) and minimum ($+0.5 A_0$). The $n=1$ contamination sample was randomly inferred from $1\%$ of the original population, and the fraction was arbitrarily scaled downward for larger $n$. Importantly, the analysis illustrates that the observed LMC overdensity ($I_c \simeq 17^{\rm m}.8$) is sampled by RRab stars contaminated by a single RCG. The broader trend toward brighter magnitudes is reproduced by increasing the number of contaminating RCGs, although contamination can stem from a diverse stellar demographic.  The pattern could be sought for in extragalactic datasets, as blended RRab and RCG stars would propel a detection $> 1^{\rm m}$ beyond uncontaminated RRab stars, and potentially a survey's faint-magnitude limit.  Moreover, modelling can be expanded to the lightcurve where blending may be evaluated in multi-parameter and Fourier space (i.e., multiband lightcurves).  Blended RRab and RCG stars would then be more confidently identified, particularly within a Wesenheit framework.  Note that the decrease in RRab magnitude as a function of amplitude relays the period-magnitude-amplitude correlation.
 
\subsection{Classical Cepheid \& RCG blends} 
\label{s-cc}
\begin{figure}
\begin{center}
\includegraphics[width=8cm]{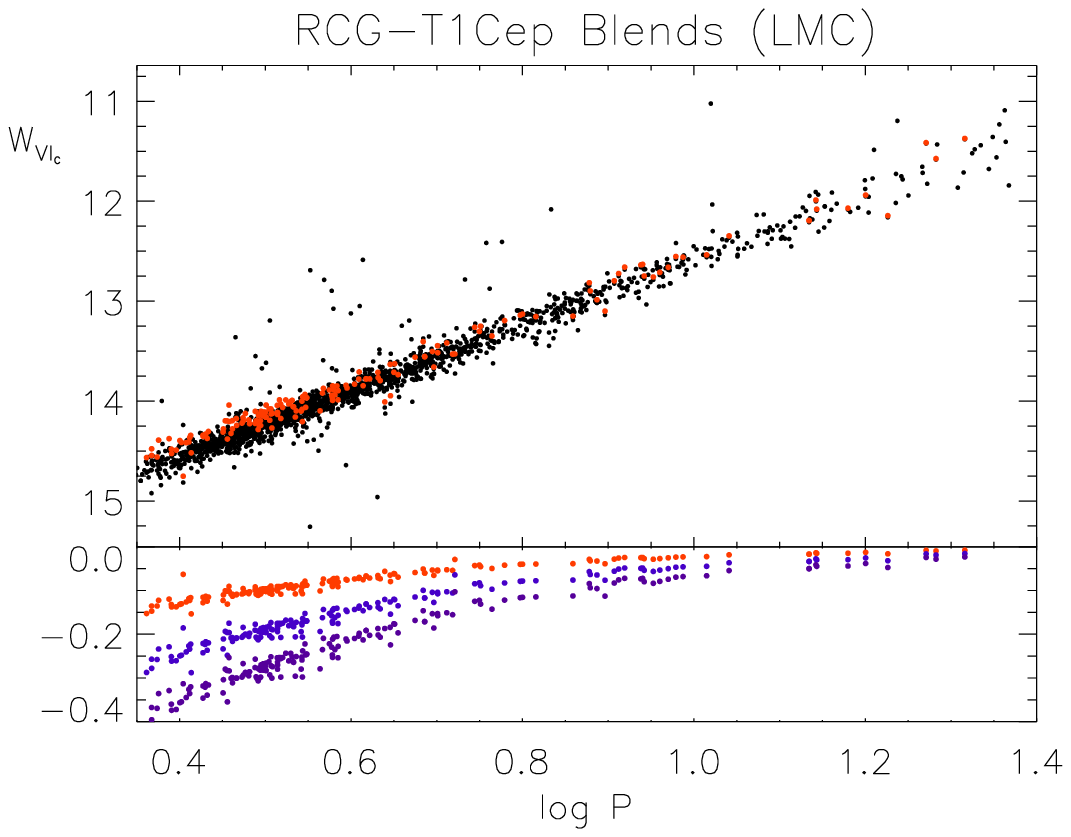} 
\includegraphics[width=8cm]{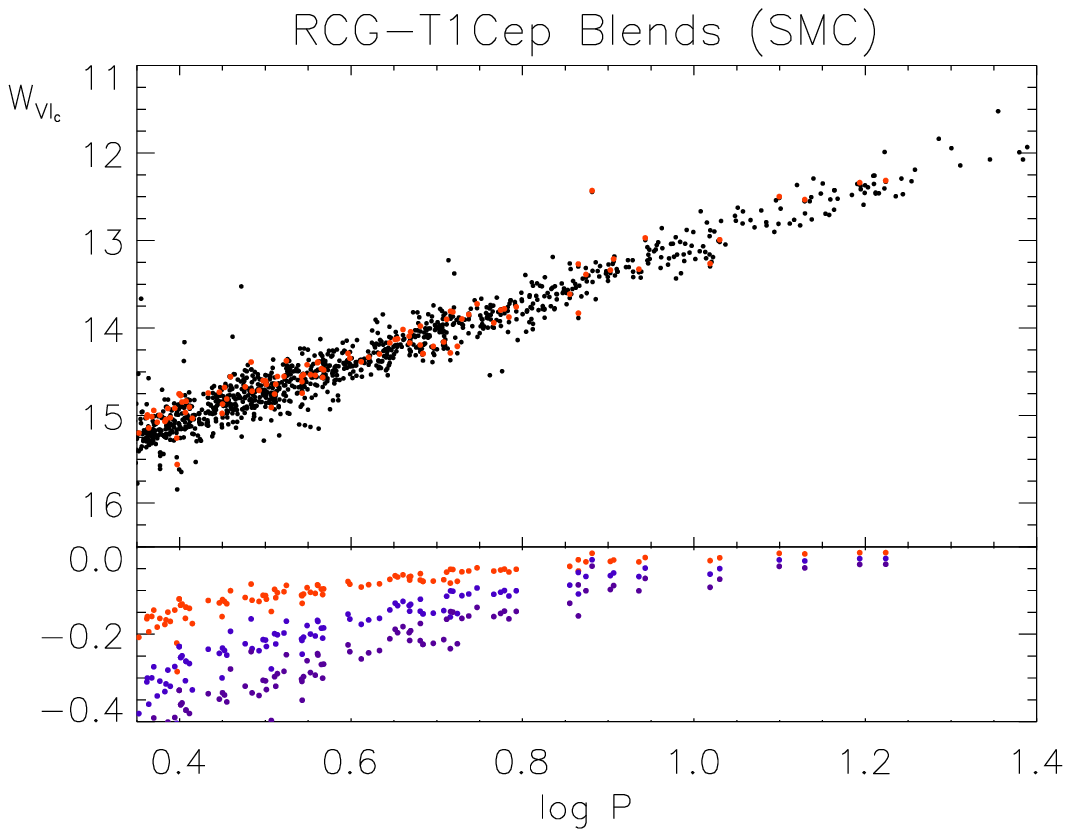} 
\caption{Simulated blends of RCGs grafted onto classical Cepheids in the Magellanic Clouds (OGLE).  The blends sway the Wesenheit ($W_{VI_c}$) slope inferred since the impact of RCGs is relatively larger for shorter-period (less-luminous) classical Cepheids.  The differentials are relayed in the bottom panels for $n=1,2,3$.}
\label{fig-2}
\end{center}
\end{figure} 

\begin{figure}
\begin{center}
\includegraphics[width=8cm]{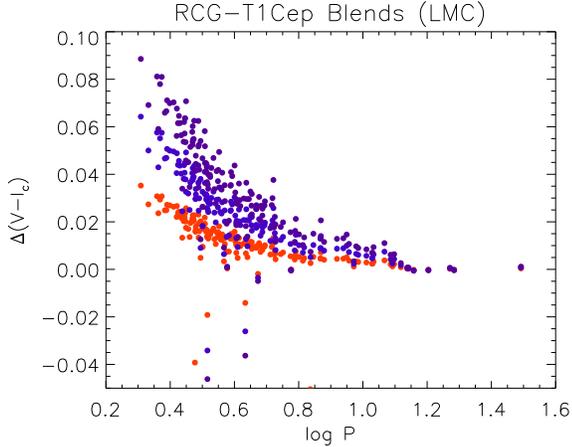} 
\caption{The impact of $n=1,2,3$ RCG blends on the color ($V-I_c$) of LMC classical Cepheids.}
\label{fig-3}
\end{center}
\end{figure}

$VI_c$ OGLE data for classical Cepheids in the Magellanic Clouds were examined \citep{so08,so10}. The blended classical Cepheid and RCG flux was evaluated assuming a mean LMC magnitude for the latter of $V_{RC}\simeq19.2$.  That was paired with the $I$-band estimate cited earlier to determine the contaminated Wesenheit magnitude.  The blended data are conveyed in Fig.~\ref{fig-2}, and represent a hypothetical remote extragalactic sample, or one sampled in a dense spiral location, or near a central region where metal-rich classical Cepheids reside.  The sample was drawn randomly from $10\%$ of the Magellanic Cloud population.  The principle objective is broadly illustrating that a shallower slope can be achieved because of the larger impact of RCGs on shorter-period classical Cepheids (less luminous).  In that instance the slope of an applied linear fit shifts from $\alpha \simeq -3.3$ to $-3.1$ (blended), and the zeropoint changes by $\simeq 0.^{\rm m}2$ ($n=1$). Yet for remote dense galaxies (e.g., NGC3370) the median blending corrections applied by the S$H_0$ES team are sizable (e.g., $>0.^{\rm m}25$), and imply that multiple RCGs could be involved in blending \citep[see also Fig.~1 in][]{su99}. The trends for $n=2,3$ are overlayed in the bottom panels of Fig.~\ref{fig-2}. The blended color $V-I_c$ changes by $+ \Delta 0.02$ for $\log{P} \simeq 0.5$ ($n=1$), and diminishes to negligible for redder longer period LMC classical Cepheids (Fig.~\ref{fig-3}).  Note that the OGLE survey of the Magellanic Clouds features numerous shorter-period classical Cepheids relative to samples of remote extragalatic variables.  The latter can be restricted to a distribution of $\log{P} > 0.8$, owing in part to the brighter luminosity (i.e., Leavitt Law) and shifts in period owing to metallicity \citep[][blue loops for $<5 M_{\sun}$]{be85}.   

NIR Wesenheit and period-reddening relations could be viable proxies for assessing whether the photometry has been accurately corrected for photometric contamination (blending) and standardized.  However, there are caveats to that assertion, one being that certain researchers advocate there is a Wesenheit zeropoint magnitude dependence ($W_{VI_c}$) on metallicity.  A brief summary is warranted, and the degeneracy with blending emerges yet again.  \citet{ma06} applied a suite of criteria aiming to mitigate blending (e.g., their Fig.~17), and subsequently argued that HST data for classical Cepheids spanning M106 (metal-rich central region to the metal-poor periphery) implied a metallicity effect of $\gamma=-0.29\pm0.16$ mag/dex ($VI_c$).  However, \citet{ma10} underscored that the revised \citet{ri09a} abundance gradient for M106 implies that the \citet{ma06} result nearly doubles to an unrealistic value \citep[see also][and \S 5 in \citealt{ma01}]{br11}, and consequently blending remained a key factor.  Yet \citet{ss11} and \citet{ge11} favored a sizable $VI_c$ metallicity dependence of $\gamma=-0.80\pm0.21$ mag/dex and $-0.62\pm0.33$ mag/dex accordingly, namely after examining HST and Large Binocular Telescope data for classical Cepheids at varying galactocentric radii in M101 and M81.  \citet{ma11} disagreed with those conclusions and advocated that photometric contamination was the contributor, since applying such an immense metallicity dependence yielded anomalous results for the distances to the comparatively nearby Magellanic Clouds (e.g., $\mu_{0,LMC} \neq 18.1$).  Indeed, \citet{ma11} reiterated that the impact of metallicity could be evaluated on the basis of several methods that did not rely exclusively on an uncertain multi-degenerate galactocentric approach, and that $W_{VI_c}$ observations of classical Cepheids were comparatively insensitive to chemical composition.  Therefore, direct empirical constraints on blending stem partially from tracking the $W_{VI_c}$ magnitude changes of classical Cepheids throughout a galaxy (i.e., from the low stellar-density periphery to near the crowded core).  Crucially, \citet{ri09b} constructed a preliminary multiband procedure to apply blending corrections, and then partly ascertained from galactocentric analyses of numerous spiral galaxies that $\gamma=-0.27\pm0.18$ mag/dex ($W_{H,VI_c}$).  That was subsequently substantially reduced by half to $\gamma=-0.14\pm0.06$ mag/dex \citep{ri16}, and for $W_{VI_c}$ the S$H_0$ES team argued for $\gamma=-0.20\pm0.05$ mag/dex. The present S$H_0$ES approach requires revision if indeed the metallicity corrections at $W_{H,VI_c}$ and $W_{VI_c}$ are relatively negligible \citep[][see also \citealt{bo08}]{ma11}.  

Lastly, the blended classical Cepheid and RCG Wesenheit magnitude (Fig.~\ref{fig-2}) was computed using an extinction law commonly employed by the OGLE team of $R_{VIc}=A_V/E(V-I_c)\simeq2.55$.  However, $R_{VIc}=2.45$ is utilized too, and the overall topic shall be discussed at length elsewhere.  There could exist both a mean extinction law offset between the calibration and target samples, and variations within each set.  For broader context, note that the Galactic Bulge sightline \textit{may} be characterized by an anomalous $R_{VI_c}$ extinction law (e.g., potentially larger dust grains) relative to the Galactic Disk \citep{ud03}. However, O-stars examined by \citet{ma16} indicate the disk along the Galactic Bulge sightlines adhere to a canonical $R_{V,BV}$, yet an anomalous visual extinction law characterizes the Carina sightline \citep{tu12b,ma16}.  The transition to given longer wavelengths mitigates that problem \citep{fr12,ma16}, but contaminating flux from RCGs (\& red evolved stars) may increase.  The Wesenheit function's nearly reddening free nature is advantageous, but a disadvantage emerges since the uncertainty stemming from the extinction law is magnified by a significant color term.  Additional research on the topic is desirable. 

\section{Conclusion}
\label{s-conclusion}
Blending was investigated as it relates to RCGs contaminating RRab stars and classical Cepheids. The former are abundant, and certain trends in OGLE RRab data were explained by contaminating the photometry with a single RCG (i.e., LMC overdensity near $I_c \simeq 17^{\rm m}.8$, Fig.~\ref{fig-l}).  The brighter blends are readily identified ($I_c$ vs.~$A_{I_c}$) in part owing to the relatively marginal magnitude offset between RRab and RCG stars.  The assertion that blending may advantageously cause variable stars to be detectable in more remote galaxies was confirmed. 
 
The larger magnitude offset between RCGs and classical Cepheids complicates the identification of that pairing.  The slope of the Wesenheit function (e.g., $W_{VI_c}$) can be valuable for identifying those blended cases and inaccurate photometry, since shorter-period classical Cepheids are acutely impacted by contamination (Fig.~\ref{fig-2}), and if a given Wesenheit relation is comparatively insensitive to metallicity \citep[][their Fig.~1]{ma11}.  Fig.~\ref{fig-2} illustrates the slope ($\alpha$) can change to lower values as a result of blending by RCGs (e.g., $| \Delta \alpha | \simeq 0.3$, $n=2$), and the RCG class provides a broader view of the contamination an evolved red demographic can impose (e.g., AGB stars, red branch giants).   

An important broader objective remains securing a reliable $H_0$ and thus further constraining cosmological models.  Investigations into the impact of blending are key to achieving that goal, as evidenced by the debate pertaining to whether the S$H_0$ES and CHP teams yielded an $H_0$ ($\simeq 74$ km s$^{-1}$ Mpc$^{-1}$) offset from the CMB framework ($H_0 = 67.4 \pm \textit{0.5}$\textbf{:} km s$^{-1}$ Mpc$^{-1}$).  The topic gained momentum in part since the authors of \citet{ts10} are no longer present to argue for $H_0 = 62.3 \pm 5$ km s$^{-1}$ Mpc$^{-1}$ \citep[see also][]{tu14}.\footnote{Admittedly, \citet{ma10} relayed that a hybrid Galactic classical Cepheid calibration tied to cluster Cepheids and HST parallaxes implied that the \citet*{sa06} classical Cepheid distances were too remote, and yielded an artificially low $H_0$ \citep[see also][]{vl07}.  However that team did not apply blending corrections \citep[e.g.,][]{ri09b} which would shift $H_0$ in the opposite direction.}  Nonetheless, \citet[][CCHP]{fr19} subsequently revised their value downward to $H_0 = 69.8 \pm 1.9$ km s$^{-1}$ Mpc$^{-1}$, hence reducing the tension relative to the CMB result.  A follow-up study by \citet[][S$H_0$ES]{yu19} argued that \citet{fr19} neglected blending and a standardization offset when determining the TRGB absolute magnitude, and consequently their $H_0$ (TRGB+SNe Ia) should be increased back upward to $72.4\pm2.0$ km s$^{-1}$ Mpc$^{-1}$, and thereby bolstering the S$H_0$ES perspective. Yet the counter effect should likewise be considered, namely the impact of blending on remote targets in all the following concurrently: the HST key project to determine $H_0$ sample \citep{gi00,fr01}, the CHP mid-IR Spitzer classical Cepheid observations \citep{fr12}, the \citet[][CCHP]{fr19} set, and the \citet{sa06} data.  That may \textit{in sum} reduce $H_0$ and supersede presently quoted uncertainties ($0.^{\rm m}04$), since separate teams did not account for blending in a similar fashion to \citet[][or \citealt{ri16}]{ri09b}. Efforts to strengthen and compare $H_0$ estimates should include an assessment of how contamination linked to remote targets is addressed. \citet{ri09b} applied significant average photometric contamination corrections ($H$-band), and \citet{ri16} note that the median blending shift tied to SN-host galaxies is $0.^{\rm m}18$. However their procedure should be independently scrutinized in concert with the systematic uncertainties flowing from the remote blending corrections applied. The HST project did not apply similar corrections for photometric contamination, but provided a sizable uncertainty and elaborate discussion tied to the phenomenon \citep[e.g.,][and references therein]{fr01}.  $H_0$ estimates could likewise be improved by inevitably pairing \textit{validated} Gaia, HIP, and HST parallaxes for classical Cepheids with a subset of cluster Cepheids where consensus exists \citep[e.g.,][and references therein]{tu12a,gr18,ri18,sh19}, and/or via a NIR Universal Wesenheit Template \citep[][and references therein]{ma11}.  The resulting analysis in tandem with other approaches may facilitate the breaking of key degeneracies, and for benchmarking the slope of the Wesenheit function, the impact of metallicity, blending, and the distance to the LMC and M106 (pending the availability of viable and independently attested blending-corrected photometry for the latter).  Indeed, multiple quantities may require adjustment that conspire to sway $H_0$ unidirectionally, and narrowing the focus may inadvertently mask a broader problem. The dawn of precision cosmology seemingly occurs in an era where a lack of agreement exists concerning fundamentals associated with the Leavitt Law, owing in part to degeneracies (e.g., blending, metallicity, and extinction law).

\acknowledgments
D.M. is grateful to the following consortia and individuals whose efforts or advice helped foster the research: OGLE, CDS, arXiv, NASA ADS, (C)CHP, S$H_0$ES, D.~Minniti, \& D.~Turner.



\end{document}